\documentclass[12pt]{article}
\usepackage{graphicx} 
\usepackage{amsmath}
\usepackage{amssymb}
\usepackage{mathrsfs}
\usepackage{amscd}
\usepackage{bbm}
\usepackage{youngtab}

\usepackage{enumerate}
\usepackage{multirow}
\usepackage{hyperref}
\usepackage[affil-it]{authblk}

\newcommand{\be}{\begin{equation}}
\newcommand{\ee}{\end{equation}}
\newcommand{\ds}{\displaystyle}

\newcommand{\iden}{\mathbbm{1}}

\newcommand{\citep}[1]{\cite{#1}}

\begin{document} 

\title{Full Current Statistics for a Disordered Open Exclusion Process}
\author{Arvind Ayyer
\thanks{arvind@math.iisc.ernet.in}}
\affil{Department of Mathematics, \\
Indian Institute of Science, \\
Bangalore - 560012, India}
\date{\today}  

\maketitle

\begin{abstract} 
We consider the nonabelian sandpile model defined on directed trees by Ayyer, Schilling, Steinberg and Thi\'ery \cite{ayyer_schilling_steinberg_thiery.sandpile.2013} and restrict it to the special case of a one-dimensional lattice of $n$ sites which has open boundaries and disordered hopping rates. 
We focus on the joint distribution of the integrated currents across each bond simultaneously, and calculate its cumulant generating function exactly. Surprisingly, the process conditioned on seeing specified currents across each bond turns out to be a renormalised version of the same process. We also remark on a duality property of the large deviation function. Lastly, all eigenvalues and both Perron eigenvectors of the tilted generator are determined.
\end{abstract}


\section{Introduction}
The computation of current fluctuations in nonequilibrium systems is an
important problem in nonequilibrium statistical mechanics. The exact
computation of the large deviation function (LDF) of the currents is difficult in
general and some overarching principles such as the additivity principle of Bodineau 
and Derrida \cite{BD1,BD3} and the macroscopic fluctuation theory of Bertini, De Sole, 
Gabrielli, Jona-Lasinio and Landim \cite{BDJGL1,BDJGL2,BDJGL3} have been established 
for driven diffusive systems. A comprehensive review of the large deviation approach 
with references to the appropriate mathematical literature is \citep{TH2013}.

These approaches are not valid for systems with ballistic transport, however, and no general principles are currently known for such systems. For some special systems, such as the ASEP (both with periodic as well as open boundary conditions), there has been considerable success using techniques from the theory of integrable systems in finding explicit formulas initially for low-order cumulants \cite{DEM1995, DM1997}, and later for the full LDF \cite{DL1998,dGE2011,LM2011}. In most cases, the LDF is given by an implicit equation. While this permits a perturbative expansion for the calculation of cumulants, global properties of the LDF are not very easy to see.

In this note, we study the total integrated current across each bond {\em simultaneously} of a disordered exclusion process on a one-dimensional lattice of length $n$ indirectly related to the well-known abelian sandpile model \citep{dhar1990}. This model is a specialisation of a directed nonabelian sandpile model on trees \cite{ayyer_schilling_steinberg_thiery.sandpile.2013}, where particles enter from the leaves, hop along the edges directed towards the root, and exit from there. Each node has a threshold. Two kinds of nonabelian sandpile models were studied there, called the {\em trickle-down sandpile model}, where one particle hops at a time, and the {\em landslide sandpile model}, where the entire block of particles at a certain node hops out and topples along the path. In this case, all thresholds are 1, and both these variants coincide.

We note that similar models have been considered in the probability literature under the name of {\em long range exclusion processes} \cite{Spitzer1970,Liggett1980}.

The steady state of this model has been shown in \cite{ayyer_schilling_steinberg_thiery.sandpile.2013} to be of product form and all 
the eigenvalues of the generator $M_n$ are known to be linear in the rates.
We will show that the tilted generator $L_n$ corresponding to joint integrated currents across all bonds has a simple structure. In particular, all its eigenvalues are known. Furthermore, we will show, using the generalised Doob transform of Chetrite and Touchette \cite{CT2015}, that the Markov process conditioned on the observed values of individual currents across each bond is a renormalised version of the same model, with the sole change being to the entry rate. This can be interpreted as a version of the additivity principle \citep{BD1}, which gives a relationship between the observed current and the density.

The plan of the article is as follows. For completeness, we will define the model in Section~\ref{sec:model} and give a summary of the results there. We will explore a ``duality'' property of the LDF in Section~\ref{sec:ldf}. The proofs, which only need basic linear algebra, will be relegated to Section~\ref{sec:proofs}. We will 
compute the Perron eigenvalue and both the corresponding eigenvectors of the tilted generator in Section~\ref{sec:tilted}. The derivation of the conditioned process via the generalised Doob transform along with all the other eigenvalues will be done in Section~\ref{sec:doob}. We conclude with some open questions in Section~\ref{sec:conc}

\section{Model definition and Statement of Results} 
\label{sec:model}

Consider a one-dimensional lattice of $n$ sites, where each site can either contain a 
single particle or can be empty. Thus, configurations can be described in terms of binary
words of length $n$, $w \in B_n := \{0,1\}^n$. The dynamics is described as follows. 
If there is a particle at site $j$, then with arbitrary rate $\alpha_j$, the particle tries to occupy site $j+1$.
If site $j+1$ already contains a particle, then it tries to move to site $j+2$, and so on, until
it finds the first vacant site $j+k$ and occupies it. If no such site exists, the particle exits the lattice
to the reservoir on the right.
That is, starting at site $j \in \{1,\dots,n\}$,
\be \label{bulktr}
\cdots\young(1\cdot\cdot\cdot10)\cdots \;\;\xrightarrow{\ds \alpha_j}\;\; 
\cdots\young(01\cdot\cdot\cdot1)\cdots
\ee
if $k \leq n-j$ and
\be \label{righttr}
\cdots\young(11\cdot\cdot\cdot\cdot\cdot1) \;\;\xrightarrow{\ds \alpha_j}\;\; 
\cdots\young(01\cdot\cdot\cdot\cdot\cdot1)
\ee
otherwise.
Particles enter from the reservoir on the left with rate $\alpha_0$ and occupy the leftmost unoccupied site,
\be \label{lefttr}
\young(1\cdot\cdot\cdot\cdot\cdot10)\cdots \;\;\xrightarrow{\ds \alpha_0}\;\; 
\young(1\cdot\cdot\cdot\cdot\cdot11)\cdots.
\ee
In the special case when all sites are occupied, the particle simply exits the system from the right,
\be \label{leftrighttr}
\young(1\cdot\cdot\cdot\cdot\cdot\cdot\cdot1) \;\;\xrightarrow{\ds \alpha_0}\;\; 
\young(1\cdot\cdot\cdot\cdot\cdot\cdot\cdot1).
\ee
While this does not change the configuration, we will need to keep track of such transitions since we are interested in the current across each bond.
In the language of directed nonabelian sandpile models \cite{ayyer_schilling_steinberg_thiery.sandpile.2013}, this corresponds to a directed tree or arborescence which is a path with all thresholds 1. The first site is a leaf, with source rate $\alpha_0$ (source rates were denoted $y_\ell$ for every leaf $\ell$ there) and the $n$'th site is the root. 

It is clear from the definition of the transitions that the steady state current across the system $j_0$ is given by the current across the leftmost bond, which is precisely the rate of transitions into the system from the left reservoir,
\be \label{sscurrent}
j_0 = \alpha_0.
\ee

We will denote by $M_n$ the Markov matrix of the system of size $n$.
For example, the Markov matrix $M_2$ in the lexicographically ordered basis 
$\{00, 01, \\ 
10, 11\}$ is given by
\be \label{eg-matrix}
M_2 = 
\left(
\begin{matrix}
-\alpha_0 & \alpha_2 & 0 & 0 \\
0 & -\alpha_0-\alpha_2 & \alpha_1 & \alpha_1 \\
\alpha_0 & 0 & -\alpha_0-\alpha_1 & \alpha_2 \\
0 & \alpha_0 & \alpha_0 & -\alpha_1-\alpha_2
\end{matrix}
\right).
\ee

The following results for this model have been shown previously.
The stationary distribution of this model is given by a product measure \cite[Theorem 2.3]{ayyer_schilling_steinberg_thiery.sandpile.2013}, where the density $\rho_j$ at site $j$ is
\be \label{density-j}
\rho_j = \frac{\alpha_0}{\alpha_0+\alpha_j}.
\ee
This holds in general for any trickle-down sandpile model. A simple consequence of this is that the 
{\em nonequilibrium partition function} $Z_n$ is given by
\[
Z_n = \prod_{j=1}^n (\alpha_0 + \alpha_j).
\]
Further, all the eigenvalues of $M_n$ are linear in the $\alpha_j$'s.
More precisely, the characteristic polynomial of $M_n$ is given by \cite[Theorem 2.10]{ayyer_schilling_steinberg_thiery.sandpile.2013}]
\be \label{charpoly}
\det(M_n - t \iden) = t 
\prod_{\emptyset \neq S \subset [n]} \left(t + \alpha_0 + \alpha_S \right),
\ee
where $[n] = \{1,\dots,n\}$ and $\alpha_S = \sum_{j \in S} \alpha_j$.
While this seems like a simple formula, no easy proofs of this result are known. In particular, the eigenvectors do not seem to have simple expressions.
For example, the eigenvalues of $M_2$ above are $0, -\alpha_0-\alpha_1, -\alpha_0-\alpha_2$ and $ -\alpha_0-\alpha_1-\alpha_2$.
In particular, all the eigenvalues are distinct, and hence the $M_n$'s are always diagonalisable. We will need this fact later. Finally, the spectral gap is given by
$\min_{i=1}^n \alpha_i$. Thus the system is gapped whenever the rates are independent of the size.

We are interested in the joint current statistics across all $n+1$ bonds in the system, which we will denote $b_0,\dots,b_n$. Note that bond $b_j$ connects sites $j$ and $j+1$ for $1 \leq j \leq n-1$, and $b_0$ (resp. $b_n$) are the bonds connecting the left (resp. right) reservoir to site 1 (resp. $n$). The standard way to analyse the current statistics across each bond is to consider the so-called {\em tilted generator}, which we will call $L_n$, in terms of parameters $\lambda_0,\dots,\lambda_n$, one for each bond. For every transition above \eqref{bulktr}--\eqref{leftrighttr}, we multiply the rate of that outgoing (but not incoming) transition by $\exp(\sum_j \lambda_j)$ where the sum is over the bonds $b_j$ crossed by the hopping particle.

For example, the tilted generator for size 2 is given by
\be \label{eg-tilted}
L_2 =  
\left( 
\begin {array}{cccc} 
-\alpha_{{0}}&\alpha_{{2}}{{\rm e}^{\lambda_{{2}}}}&0&0\\
\noalign{\medskip}0&-\alpha_{{2}}-\alpha_{{0}}&\alpha_{{1}}{{\rm e}^{\lambda_{{1}}}}&\alpha_{{1}}{{\rm e}^{\lambda_{{1}}+\lambda_{{2}}}}\\
\noalign{\medskip}\alpha_{{0}}{{\rm e}^{\lambda_{{0}}}}&0&-\alpha_{{1}}-\alpha_{{0}}&\alpha_{{2}}{{\rm e}^{\lambda_{{2}}}}\\
\noalign{\medskip}0&\alpha_{{0}}{{\rm e}^{\lambda_{{0}}}}&\alpha_{{0}}{{\rm e}^{\lambda_{{0}}+\lambda_{{1}}}}&-\alpha_{{2}}-\alpha_{{1}}+\alpha_{{0}} \left( {{\rm e}^{\lambda_{{0}}+\lambda_{{1}}+\lambda_{{2}}}}-1 \right) \end {array}
\right).
\ee
Note that the term $\alpha_{{0}} {{\rm e}^{ \lambda_{{0}} + \lambda_{{1}} + \lambda_{{2}}}}$ in the $(2,2)$ entry of $L_2$ corresponds to the transition \eqref{leftrighttr}.

It is well-known that the largest eigenvalue of $L_n$ is the joint cumulant generating function of the currents $\mu(\lambda_0,\dots,\lambda_n)$. We will show  in Section~\ref{sec:tilted} that 
\be \label{jointcumulantgf}
\mu_n \equiv \mu(\lambda_0,\dots,\lambda_n) = \alpha_0 \left( \exp \left(\sum_{j=0}^n \lambda_j \right) -1 \right),
\ee
This has the structure of the cumulant generating function of a Poisson distribution. While it is clear that the current across $b_0$ has that form, it is not obvious apriori that the joint distribution will also be Poissonian.
Surprisingly, it will turn out that all the other eigenvalues of $L_n$ are the same as those of $M_n$, so that the characteristic polynomial of $L_n$ is given by 
\be \label{tilted-charpoly}
\det(L_n - t \iden) = \left(t - \mu_n \right)
\prod_{\emptyset \neq S \subset [n]} \left(t + \alpha_0 + \alpha_S \right).
\ee
These results will be proved along with formulas for both Perron eigenvectors in Section~\ref{sec:tilted}, 

The fact that $\mu_n$ depends only on the sum of the $\lambda_j$'s is expected to hold generically in systems without a condensation phase transition (unlike \cite{HRS2005}). In particular, it should hold for any nontrivial ergodic finite-state exclusion process.
For example, for the usual open one-dimensional TASEP \cite{DEHP93} with $n=2$, the tilted generator in the lexicographically ordered basis for the joint currents analogous to \eqref{eg-tilted} is given by
\[
\tilde{L}_2 = 
\left(
\begin {array}{cccc} 
-\alpha&\beta{{\rm e}^{\lambda_{{2}}}}&0&0\\
\noalign{\medskip}0&-\alpha-\beta&{{\rm e}^{\lambda_{{1}}}}&0\\
\noalign{\medskip}\alpha{{\rm e}^{\lambda_{{0}}}}&0&-1&\beta{{\rm e}^{\lambda_{{2}}}}\\
\noalign{\medskip}0&\alpha{{\rm e}^{\lambda_{{0}}}}&0&-\beta
\end {array}
\right).
\]
There's no simple expression for the largest eigenvalue of $\tilde{L}_2$, but
the characteristic polynomial $\det(\tilde{L}_2 - t \iden)$ has the expression
\[
\left( t+1 \right)  \left( t+\beta \right)  \left( t+\alpha \right)  \left( \alpha+t+\beta \right) - \alpha\,\beta\, {{\rm e}^{\lambda_{{0}}+\lambda_{{2}}+\lambda_{{2}}}}\left( 2\,t+\alpha+\beta \right),
\]
which again shows that largest eigenvalue, which is the joint cumulant generating function for the currents across individual bonds, depends only on $\lambda_{{0}}+\lambda_{{2}}+\lambda_{{2}}$. 

Let $Q_t^{(0)},\dots,Q_t^{(n)}$ be the number of particles hopping across bonds $b_0,\dots,b_n$ upto time $t$ starting from a certain initial configuration.
For such a generic nonequilibrium system, we expect the joint distribution for large times to be given by the large deviation function $F(j)$ according to
\be \label{joint-current-dist}
\mathbb{P} \left(\frac{Q_t^{(0)}}t = j_0,\dots,\frac{Q_t^{(n)}}t = j_n \right)  \overset{t \to \infty}{\sim } \;\prod_{i=1}^n \delta(j_0-j_i) \;
{\rm e}^{-t F(j_0)}.
\ee
Note that the currents across individual bonds coincide in the large time limit.
Thus, the large deviation function of the current is a function of a single variable.
This is corroborated by the fact that the cumulant generating function $\mu_n$ in \eqref{jointcumulantgf} depends only on the sum of its arguments. 
The large deviation function $F$ is given by the Legendre transform of $\mu_n$, which is easily seen to be
\be \label{largedeviationfn}
F(j) = j \log \left( \frac{j}{\alpha_0} \right) - j + \alpha_0.
\ee
Curiously, this does not depend on the size of the system. This formula appears often, for instance in open systems like ours, where the entry rate of particles is independent of the configuration.

Recently, Chetrite and Touchette have investigated Markov processes conditioned on rare events \cite{CT2015}. The idea there is that, if one sees a large fluctuation in an observable in a Markov process, then one can obtain a natural Markov process from the original one, where this fluctuation corresponds to the stationary value of this observable. Physically, one can think of the stationary distribution of the modified process as a ``quasistationary'' distribution of the original process when the fluctuation lasts for a long time. The main mathematical ingredient there is a generalisation of the Doob transform. 

Consider the Markov process conditioned on having currents $j_0,\dots,j_n$ across bonds $b_0,\dots,b_n$ specified by the conjugate variables $\lambda_0,\dots,\lambda_n$ as above. Because of the structure of the cumulant generating function \eqref{jointcumulantgf},
each $j_i$ is forced to be equal to $\alpha'_0 = \alpha_0 \exp(\lambda_0 + \dots + \lambda_n)$. 

We expect our conditioned Markov process to have $\alpha'_0$ as the entry rate. In fact, we will show, in Section~\ref{sec:doob}, that the new process is nothing
but a {\em renormalised} version of the original process with this being the sole change. We will do so by computing the Markov matrix $M'_n$ of the conditioned process. For example, the conditioned Markov matrix $M'_2$ is
given by \eqref{eg-matrix} with $\alpha_0$ replaced by $\alpha'_0$.
An immediate consequence of this is that the quasistationary distribution for completely arbitrary current fluctuations is also a product distribution with density $\rho'_j$ at site $j$ given, using \eqref{density-j}, by
\be \label{cond-density-j}
\rho'_j = \frac{\alpha'_0}{\alpha'_0 + \alpha_j}.
\ee
This gives a nontrivial relationship between the empirical currents and empirical densities. For instance, consider the case when all hopping rates $\alpha_j$'s are equal to 1. The density in the steady state in that case is uniformly $1/2$ throughout the system and the current is 1. If the observed current is $j ( = \alpha'_0)$, then we see that the corresponding quasistationary distribution, using \eqref{cond-density-j}, has density $j/(1+j)$ again uniformly throughout the system. This density-current relationship might be an indication of a more general phenomenon.

It also follows immediately that the characteristic polynomial of the conditioned Markov matrix $M'_n$ is given by
\be \label{cond-charpoly}
\det(M'_n - t \iden) = t \prod_{\emptyset \neq S \subset [n]} 
\left(t + \alpha'_0 + \alpha_S \right).
\ee

\section{Large deviation function}
\label{sec:ldf}
The large deviation function $F(j)$ is generically a convex function with a minimum being zero at the steady state value $j_0$. Therefore, for small deviations of $j$ from $j_0$ with $j>j_0$ (say), we expect there to be a ``dual'' current $j'<j_0$ such that $F(j) = F(j')$.  Note that this argument is quite general and does not have any apriori connection with other properties of the LDF, such as the Gallavotti-Cohen symmetry \cite{GC1995}.
In particular, these dual currents are also present in totally asymmetric dynamical systems like the TASEP and the model considered here. Although the LDF of the TASEP has been explicitly computed in \cite{DL1998}, it is given by an implicit equation, and it is not easy to see how to compute the dual current. 

In our case, $F(j)$ is given by \eqref{largedeviationfn}, with the steady state value given by \eqref{sscurrent}. Without loss of generality, we can set $\alpha_0 = 1/{\text e}$ by rescaling time to obtain
\[
F(j) = j \log j + 1/{\text e}.
\]
Suppose $0 < j_1 < 1/{\text e}$ and $1/{\text e} < j_2 < 1$. Then the solution to $F(j_1) = F(j_2)$ is given by
\[
j_2 = {\text e}^{W(j_1 \log j_1)},
\]
where $W$ is the Lambert $W$-function, which is the inverse function of $x \exp(x)$.
One can check that $j_2$ monotonically decreases as $j_1$ increases, as expected.

It might be worth understanding the relationship between the two values of the current with the same LDF in other models where the LDF is known, such as the TASEP on a ring \citep{DL1998}, the ASEP with open boundary conditions \cite{LM2011} and the open TAZRP \cite{HRS2005}.

\section{Proofs of the results} 
\label{sec:proofs}
We give the details of the proofs here for the interested reader. 
In Section~\ref{sec:tilted}, we will prove the formulas for the largest eigenvalue  \eqref{jointcumulantgf} of the tilted generator $L_n$ and the corresponding left and right eigenvectors. In Section~\ref{sec:doob}, we will show that the conditioned Markov matrix $M'_n$ has the same structure as the original Markov matrix $M_n$, which will prove \eqref{cond-density-j} and \eqref{cond-charpoly}. This will then be used to show that the eigenvalues of $L_n$ can be determined from \eqref{tilted-charpoly}.
Most of the ideas involve basic linear algebra.

\subsection{Tilted generator}
\label{sec:tilted}
The generic nonzero off-diagonal element of the tilted generator $L_n(w',w)$ for binary words $w,w' \in B_n$ is the rate of the transition from $w$ to $w'$ times exponential factors for each bond hopped in the transition. If the particle from site $i$ of $w$ ends up in site $j>i$ in $w'$, then
\be \label{Ln-entries}
L_n(w',w) = \alpha_i \exp(\lambda_i+\dots+\lambda_{j-1}).
\ee
One can check that this includes the case of particles entering from the left ($i=0$)
and the case when the particles exit from the right ($j=n+1$). The diagonal elements of $L_n$ are identical to those of $M_n$ except in the special case when $w=w'=(1,\dots,1)$, where
\be \label{Ln-last-entry}
L_n(w,w) = M_n(w,w) + \mu_n,
\ee
corresponding to transition \eqref{leftrighttr}.

We will prove the formula for the Perron eigenvalue \eqref{jointcumulantgf} by explicitly showing that the row vector $\langle \ell_n |$
whose entries are given by
\be \label{left-eigenvec-Ln}
\ell_n(w_1,\dots,w_n) = \prod_{\substack{i=1 \\ w_i = 1}}^n \exp(\lambda_i + \cdots + \lambda_n) = \prod_{i=1}^n \exp(\lambda_i \times \#
\{\text{1's in $w_1,\dots,w_i$}\})
\ee
is a left eigenvector of $L_n$. To do so, we will show that the master equation $\langle \ell_n | L_n = \mu_n \langle \ell_n |$ in coordinate form
\be \label{master-eq-Ln}
\sum_{w' \neq w} \ell_n(w') L_n(w',w) + \ell_n(w) L_n(w,w) = \mu_n \ell_n(w)
\ee
is satisfied. Let $n_1(w)$ be the number of 1's in $w$. The number of transitions from $w$ is clearly $n_1(w)+1$. First, consider the transitions corresponding to the movement of one of the $n_1(w)$ 1's in $w$. Suppose $w_i = 1$ and the transition with rate $\alpha_i$ takes this particle to site $j$ in $w'$. 
Then $L_n(w',w) = \alpha_i \exp(\lambda_i+\dots+\lambda_{j-1})$, as argued before. Now, using \eqref{left-eigenvec-Ln}, we see that
\be \label{bulk-trans}
\frac{\ell_n(w')}{\ell_n(w)} = \frac{1}{\exp(\lambda_i+\dots+\lambda_{j-1})}.
\ee
Thus, $\ell_n(w') L_n(w',w) = \alpha_i \ell_n(w)$. But since $\alpha_i$ is precisely the rate of this transition, this exactly cancels the diagonal term in $L_n(w,w)$. This argument also goes through when the particle exits the system from the right, $j=n+1$.

Now, consider the transition for the entry of the particle from the left, namely $i=0$. Suppose the particle ends up at site $j$ in $w'$. Then $L_n(w',w) = \alpha'_0 \exp(\lambda_0+\dots+\lambda_{j-1})$. Again, using \eqref{left-eigenvec-Ln},
\be \label{boundary-trans}
\frac{\ell_n(w')}{\ell_n(w)} = \exp(\lambda_j+\dots+\lambda_{n}),
\ee
which gives $\ell_n(w') L_n(w',w) = \alpha'_0 \ell_n(w)$. This time, the outgoing rate is $\alpha_0$, which implies $L_n(w,w)$ contains the term $-\alpha_0 \ell_n(w)$. 
These two terms combine to give $\mu_n \ell_n(w)$, as desired.
The special case of $j=n+1$ is again similar. Combining these factors, we have verified the master equation \eqref{master-eq-Ln} in all cases. 

Now, since $L_n$ is an irreducible matrix with off-diagonal nonnegative entries, we can apply the Perron-Frobenius theorem. Thus, $L_n$ has a unique maximal eigenvalue (called the Perron eigenvalue) whose eigenvector entries can be chosen to be positive simultaneously. Further, all other eigenvectors necessarily have entries which are both positive and negative. Since we have constructed an explicit positive left-eigenvector $\langle \ell_n |$ \eqref{left-eigenvec-Ln} of $L_n$, $\mu_n$ is the Perron eigenvalue. This proves that $\mu_n$ is given by \eqref{jointcumulantgf}.

A curious observation is that $\langle \ell_n|$ given by \eqref{left-eigenvec-Ln}, when normalised to be a probability distribution, is of product form, with the normalisation factor given by
\[
\prod_{j=1}^n \left(1 + \exp(\lambda_j + \dots + \lambda_n) \right).
\]
It turns out that with very little extra effort and the same strategy we can show that the right Perron eigenvector $|r_n \rangle$ of $L_n$, appropriately normalised, is also a product measure with density at the $j$'th site given by
\[
\frac{\alpha_0 \exp(\lambda_0 + \cdots + \lambda_{j-1})}{\alpha_0 \exp(\lambda_0 + \cdots + \lambda_{j-1}) + \alpha_j}.
\]
We leave the details for the interested reader.
We will prove the formula \eqref{tilted-charpoly} for the other eigenvalues of $L_n$ in Section~\ref{sec:doob}.

\subsection{Generalised Doob Transform}
\label{sec:doob}
For a finite state continuous-time Markov chain, the generalised Doob transform \cite{CT2015} of $L_n$ has Markov matrix $M'_n$ whose entries are given by
\be \label{cond-generator}
M'_n(w',w) = L_n(w',w) \frac{\ell_n(w')}{\ell_n(w)} - \delta_{w',w}\mu_n.
\ee
One can check that $\sum_{w'} M'_n(w',w) = 0$. The entries of $M'_n$ are easily determined from the analysis of the tilted generator in Section~\ref{sec:tilted}. The entries of $L_n$ are given by \eqref{Ln-entries} and we can substitute them in \eqref{cond-generator}.
For the off-diagonal elements, whenever $L_n(w',w)$ is nonzero because of the movement of a particle in the bulk in $w$, we get using \eqref{bulk-trans},
\[
M'_n(w',w) = \alpha_i =  M_n(w',w),
\]
and whenever $L_n(w',w)$ is nonzero because of the entry of a particle from the left reservoir, \eqref{boundary-trans} leads to
\[
M'_n(w',w) = \alpha'_0  = M_n(w',w) \exp(\lambda_0+\dots+\lambda_{n}).
\]
Since the diagonal elements of $L_n$ are the same as that of $M_n$ (except for the fully occupied state) and each contains the term $-\alpha_0$, subtracting $\mu_n$ from these entries will just replace $\alpha_0$ by $\alpha'_0$. When $w$ corresponds to the fully occupied state,\eqref{Ln-last-entry} gives $M'_n(w,w) = M_n(w',w)$. 

To summarise, the nonzero entries of $M'_n$ coincide with those of $M_n$ unless the latter are $\alpha_0$, in which case they get replaced by $\alpha'_0$. Thus, the generalised Doob transformed Markov process is nothing but a renormalised version of the original process, where the entry rate of particles from the left reservoir is replaced by $\alpha'_0$. 

From this, it follows that the stationary distribution of the transformed process has a product measure with density $\rho'_j$ at site $j$ given by \eqref{cond-density-j}. The eigenvalues of the conditioned Markov matrix \eqref{cond-generator} are also clearly read off from \eqref{cond-charpoly}.

We can use the latter fact to determine all the eigenvalues of $L_n$ in the standard way \cite{CT2015}. Suppose $a_n$ is a left eigenvector of $M'_n$ with eigenvalue $c_n$. Using \eqref{cond-generator}, one can check that the vector with entries given by $(\ell_n(w) a_n(w))_{w \in B_n}$ is a left eigenvector of $L_n$ with eigenvalue $c_n+\mu_n$. This confirms that the Perron eigenvalue of $L_n$, which corresponds to $c_n=0$ is $\mu_n$. Moreover, from \eqref{cond-charpoly}, we see that all other eigenvalues of $M'_n$ contain a term
$-\alpha'_0$ and on adding $\mu_n$, this term gets replaced by $-\alpha_0$. Since all the eigenvalues of $M'_n$ are distinct, this accounts for all eigenvalues of $L_n$ and proves \eqref{tilted-charpoly}.

\section{Conclusions}
\label{sec:conc}

In this article we have calculated the joint current cumulant generating function across each bond in a disordered totally asymmetric long range exclusion process. Further, we have established, using the generalised Doob transform, a relationship between observed atypical currents across each of the bonds and the expected local density in the system. We also studied properties of the large deviation function in the long time limit, i.e. close to stationarity. We note that the model does not possess local detailed balance, because of which we cannot even hope to test recently discovered tenets such as the Gallavotti-Cohen symmetry \cite{GC1995}, the additivity principle \citep{BD1} or the macroscopic fluctuation theory \citep{BDJGL1}. However, we have shown how dual currents are related in this model.

It would be of interest to generalise this model to include hopping in both directions such that the full current statistics are still given as explicitly as the cumulant generating function $\mu_n$ \eqref{jointcumulantgf} here. This would allow us to see how local detailed balance is related to the the dual current. If we have such an explicit formula, we could also investigate other properties that the LDF should satisfy.

A natural question would be the calculation of the joint large deviation function in finite time, which would depend on $j_0,\dots,j_n$ in a nontrivial way. 
This would involve calculating the eigenvectors of the tilted generator $L_n$. While the eigenvalues have been determined here in \eqref{tilted-charpoly}, these are calculated using the sophisticated machinery of $\mathscr{R}$-trivial monoids \cite{ayyer_schilling_steinberg_thiery.2013, ayyer_schilling_steinberg_thiery.sandpile.2013}, and we do not know of any simple way of deriving them using, for instance, standard linear algebraic techniques. This suggests that calculating the eigenvectors is a nontrivial exercise, even in the case that all $\alpha_j$'s are equal to 1.

Finally, we note that the fact that $\mu_n$ depends only on $\sum_{j} \lambda_j$ is related to there being a single current and no condensation phase transition a la \cite{HRS2005}. As far as we know, there is no mathematical proof of this result. A proof of such a result might lead to a better understanding of current correlations.

\section*{Acknowledgements}
We thank R. Chetrite, A. Dhar, K. Mallick, G. Sch\"utz and H. Touchette for discussions.
We would like to acknowledge support in part by a UGC Centre for Advanced Study grant.

\bibliographystyle{alpha}
\bibliography{sandpile}

\end{document}